\documentclass[%
  english,
  reprint,
  superscriptaddress,
  amsmath,amssymb,
  aps,
  prb,
  floatfix,
]{revtex4-2}

\usepackage{graphicx}% Include figure files
\usepackage[tight]{subfigure}
\usepackage{dcolumn}% Align table columns on decimal point
\usepackage{bm}% bold math
\usepackage[colorlinks,linkcolor={blue},citecolor={blue},urlcolor={blue}]{hyperref}% add hypertext capabilities
\usepackage{multirow}
\usepackage{array}
\usepackage{booktabs}
\usepackage{ctable}
\usepackage{upgreek}
\usepackage{epsfig}
\usepackage{mathrsfs}
\usepackage{amssymb}
\usepackage{amsbsy}
\usepackage{color}
\usepackage{cancel}
\usepackage{pifont}
\usepackage{marginnote}
\usepackage{dsfont}
\usepackage{babel}
\newcommand{\abs}[1]{\lvert#1\rvert}

\newcommand{\bcen}{\begin{center}}
\newcommand{\ecen}{\end{center}}
\newcommand{\btab}{\begin{tabular}}
\newcommand{\etab}{\end{tabular}}
\newcommand{\bdes}{\begin{description}}
\newcommand{\edes}{\end{description}}

\newcommand{\beq}{\begin{equation}}
\newcommand{\eeq}{\end{equation}}
\newcommand{\bea}{\begin{eqnarray}}
\newcommand{\eea}{\end{eqnarray}}

\newcommand{\bary}{\begin{array}}
\newcommand{\eary}{\end{array}}
\newcommand{\benum}{\begin{enumerate}}
\newcommand{\eenum}{\end{enumerate}}
\newcommand{\bitem}{\begin{itemize}}
\newcommand{\eitem}{\end{itemize}}

\newcommand{\mean}[1]{\langle #1 \rangle}

\newcommand{\Eqn}[1] {Eq.~(\ref{#1})}

\newcommand{\Fig}[1]{Fig.~\ref{#1}}
\newcommand{\Figg}[1]{Figure~\ref{#1}}
\newcommand{\Figm}[1]{~\ref{#1}}
\newcommand{\Figs}[1]{Figs.~\ref{#1}}

\newcommand{\opt}[1]{}

\newcommand{\titlename}{Finite temperature study of correlations
in bilayer band-insulator}

\begin{document}

% Title of paper

\title{\titlename}

\author{Yogeshwar Prasad}
\email{yogeshwar@iisc.ac.in}

\affiliation{Center for Condensed Matter Theory,
  Department of Physics, Indian Institute of Science,
  Bangalore 560012, India}

\date{\today}

\begin{abstract}

  We perform the finite-temperature determinant quantum Monte Carlo
simulation for the attractive Hubbard model on the half-filled
bilayer square lattice. Recent progress on optical lattice
experiments lead us to investigate various single-particle properties
such as momentum distribution and double occupancies which should be
easily measured in cold-atom experiments. The pair-pair and the
density-density correlations have been studied in detail, and through
finite-size scaling, we show that there is no competing charge
density wave order in the bilayer band-insulator model and that the
superfluid phase is the stable phase for the interaction range
$\abs{U}/t = 5-10$. We show the existence of two energy scales in the
system as we increase the attractive interaction, one governing the
phase coherence and the other one corresponding to the molecule
formation. In the end, we map out the full $T-U$ phase diagram and
compare the $T_c$ obtained through the mean-field analysis. We
observe that the maximum $T_c/t(= 0.27)$ occurs for $\abs{U}/t = 6$,
which is roughly twice the reported $T_c$ of the single-layer
attractive Hubbard model.

\end{abstract}

% insert suggested PACS numbers in braces on next line
%\pacs{71.10.Fd, 37.10.Jk,74.78.Fk, 74.78.Na}

\maketitle

\section{Introduction}

  The half-filled attractive Hubbard model (AHM), at low
temperatures, shows the $s$-wave superfluidity with a BEC-BCS
crossover along with the charge density wave (or pair density wave)
\cite{Nozieres1985}. In the weak-coupling regime $(U << t)$, fermions
of opposite spin and same momentum state form loosely bound Cooper
pairs and go to the BCS state at the critical temperature $T_c$,
which increases with $\abs{U}/t$. As we go towards the
strong-coupling regime $(U >> t)$, the strong interactions between
the particles lead to the formation of the bound pairs that condense
to form a BEC state at $T_c$, which decreases as $t^2/\abs{U}$. The
fermionic pairs in the BEC state can be regarded as the hard-core
bosons, and the tunneling of the pairs is dominated by the
second-order tunneling $t^2/\abs{U}$.

  The advancement in the field of cold atoms has generated a lot of
interest in the condensed matter community due to a control over
various parameters such as tuning the interaction between particles
by the Feshbach resonance, tuning the hopping between lattice sites
by laser intensity, and so on. The lowest temperature that has been
achieved so far in experiments on the AHM is $T \sim 0.4t$
\cite{Mitra2018}. Earlier theoretical work on half-filled AHM has
shown $T_c \sim 0.13t$ for $\abs{U}/t = 8$ \cite{Scalettar1989},
whereas the maximum $T_c \sim 0.17t$ at density $n = 0.7$ and
$\abs{U}/t = 4$ has been reported \cite{Paiva2010, Fontenele2022}.
The process to lower the temperature experimentally in optical
lattice systems has been hampered by the cooling problem (entropy
issues) \cite{Ho2009A,Ho2009B,Bernier2009,Paiva2010,McKay2011}.
\vspace{-4mm}
\begin{figure}[b]
  \centering
  \centerline{
  \includegraphics[width=0.58\columnwidth]{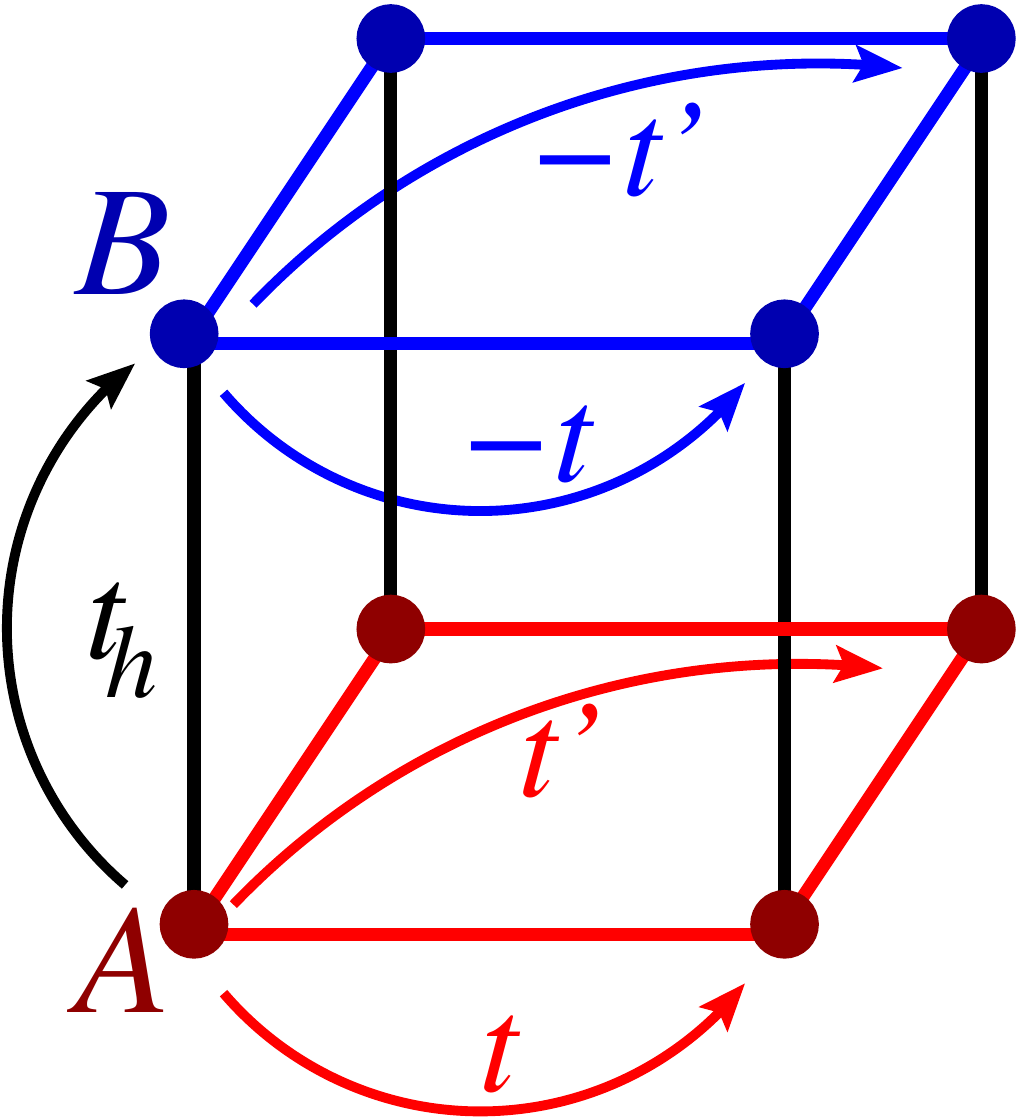}
  \includegraphics[width=0.4\columnwidth]{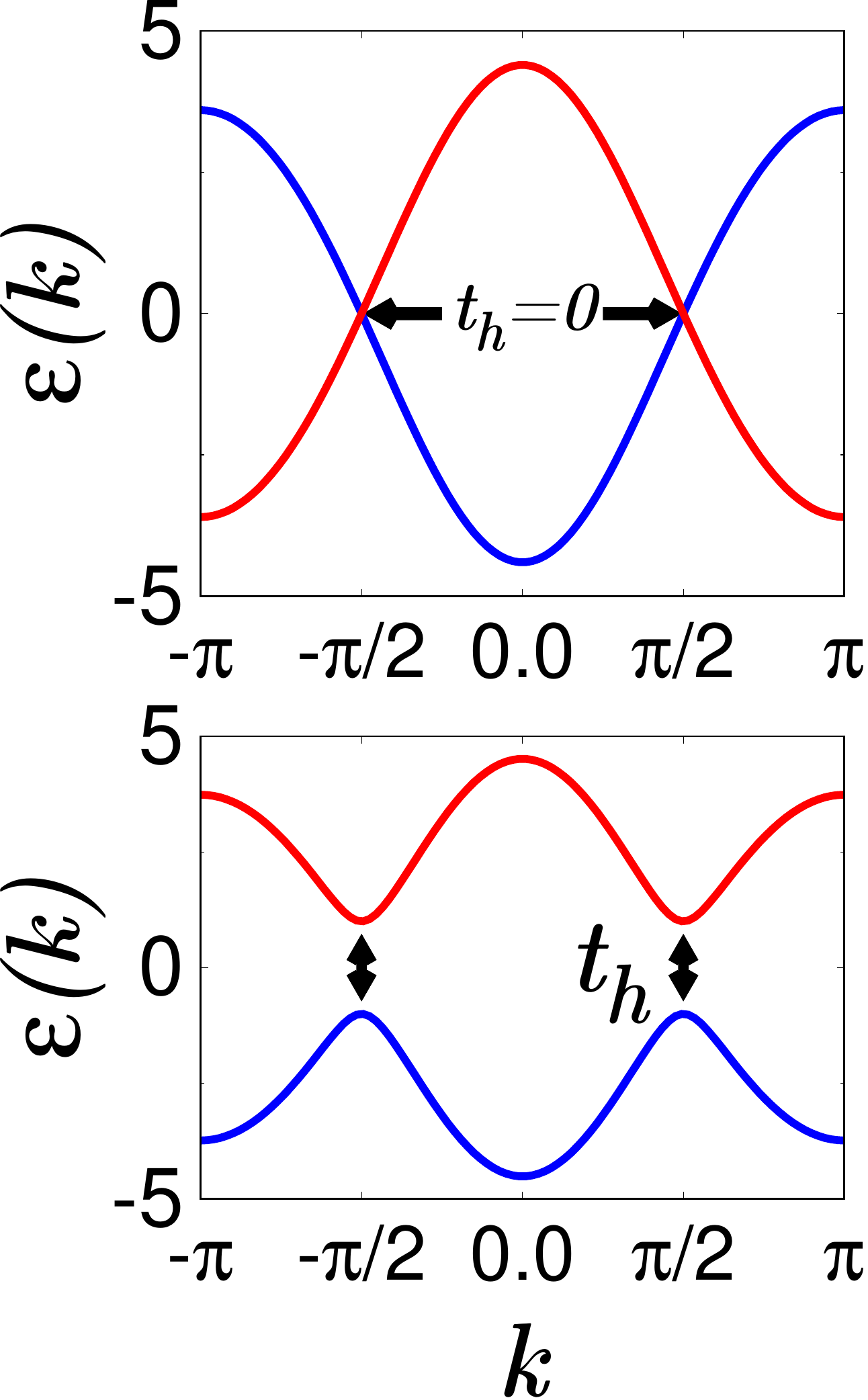}
  }
  \caption{Schematic of the bilayer band insulator. Nearest-neighbor
    hopping $t$ and next-nearest-neighbor hopping $t'$ in layer $A$
    is opposite to that of layer $B$ of the square lattice. Both
    layers have been hybridized by hopping $t_h$. The right panel
    shows the energy dispersion for $t_h = 0$ and $t_h \neq 0$.}
  \label{fig:scheme}
\vspace{-4mm}
\end{figure}

  Recent progress made in the experiments towards the realization of
the attractive Hubbard model has been of great interest
\cite{Mitra2018,Gall2020}. There have been some theoretical attempts
to ``increase" the characteristic temperature $T_c$ which can be
achieved experimentally \cite{Prasad2014,Haldar2014}. In this work,
we focus on the bilayer attractive Hubbard model band-insulator model
discussed in Ref. \cite{Prasad2014}. The model is a bilayer square
lattice model, as shown in \Fig{fig:scheme}. Both layers have been
hybridized by the coupling $t_h$. The hoppings in both layers were
taken to be opposite to each other such that the in-plane energy
dispersions in the two layers are of the {\it opposite} signs, i.e.,
$\epsilon_A({\bf k})=-\epsilon_B({\bf k}) \equiv \epsilon({\bf k})$.
The idea is to start with a low entropy state and explore the
possibility of realizing a superfluid. At half filling, for a finite
$t_h$ and for small values of $\abs{U}$, the system is in the normal
band-insulator state. With the increase in the on-site attractive
interaction $\abs{U}$, a quantum phase transition occurs at
$\abs{U_c}$, ushering in a superfluid state. Detailed analysis
including Gaussian fluctuations and variational Monte Carlo (VMC)
calculations establish that there are no competing orders such as an
intervening charge density wave (CDW) and confirm that the
superfluid state is stable at $T = 0$ \cite{Prasad2014}.
Band-insulator-superfluid transitions have been studied earlier in
other contexts \cite{Kohmoto1990,Nozieres1999}. The quantum Monte
Carlo studies for the attractive Hubbard model have been done
in the past for single layer \cite{Moreo1991,Santos1992,Kyung2001,
Paiva2004} as well as bilayer \cite{Zujev2014} square lattices. These
simulations motivated us to analyze the bilayer band-insulator model
using the determinant quantum Monte Carlo (DQMC) technique.

  We use the DQMC technique \cite{Blankenbecler1981,Santos2003} to
study the equilibrium properties of the bilayer band-insulator model
with attractive Hubbard interaction. A key point with this model is
that there is no sign problem for the attractive on-site interaction.
So it is possible to explore the physics of this bilayer model at low
temperatures. We have examined the nature of the pairing correlations
for our model through DQMC, which provides an approximation-free
solution of the discussed model. In the end, we perform the scaling
analysis and estimate $T_c$ through the finite-size scaling $(FSS)$.

\vspace{-4mm}
\section{Model and Computational Method}
\vspace{-2mm}

  We start with the Hamiltonian of the bilayer square lattice model,
$\mathcal{H} = \mathcal{H}_K + \mathcal{H}_U$, where
\vspace{-3mm}
\begin{eqnarray}
  \label{eqn:bilayer_Hamiltonian}
  \nonumber
  \mathcal{H}_K \hspace{1mm} = && \hspace{1mm}
  \overbrace{- \hspace{1mm} t \sum_{<{\bf ij}>,\sigma}
  (a^\dagger_{{\bf i}\sigma} a_{{\bf j}\sigma} + h.c.)
  - t' \sum_{<{\bf ii'}>,\sigma}
  (a^\dagger_{{\bf i}\sigma} a_{{\bf i'}\sigma}
  + h.c.)}^{\mbox{\scriptsize{$A$-layer}}} \\
  \nonumber
  &&
  \overbrace{+ \hspace{1mm} t \sum_{<{\bf ij}>,\sigma}
  (b^\dagger_{{\bf i}\sigma} b_{{\bf j}\sigma} + h.c.)
  + t' \sum_{<{\bf ii'}>,\sigma}
  (b^\dagger_{{\bf i}\sigma} b_{{\bf i'}\sigma}
  + h.c.)}^{\mbox{\scriptsize{$B$-layer}}} \\
  \nonumber
  && \underbrace{- \hspace{1mm} \sum_{{\bf i},\sigma}
  t_h({\bf i}) (a^\dagger_{{\bf i}\sigma} b_{{\bf i}\sigma}
  + h.c.)}_{\mbox{\scriptsize{$A$-$B$ Layer hybridization}}}
  - \hspace{1mm} \mu \sum_{{\bf i},\sigma}
  (a^\dagger_{{\bf i}\sigma} a_{{\bf i}\sigma}
  + b^\dagger_{{\bf i}\sigma} b_{{\bf i}\sigma}); \\
  \mathcal{H}_U \hspace{1mm} = && \hspace{1mm}
  \underbrace{- \hspace{1mm} U \sum_{\bf i}
  (a^\dagger_{{\bf i}\uparrow} a^\dagger_{{\bf i}\downarrow}
  a_{{\bf i}\downarrow} a_{{\bf i}\uparrow}
  + b^\dagger_{{\bf i}\uparrow} b^\dagger_{{\bf i}\downarrow}
  b_{{\bf i}\downarrow} b_{{\bf i}\uparrow})
  }_{\mbox{\scriptsize{Interaction \hspace{1mm} Hamiltonian}}}
\vspace{-3mm}
\end{eqnarray}
in the presence of the on-site Hubbard interaction (``$-U$"). The
first term $\mathcal{H}_K$ represents the hopping (kinetic energy) of
the fermions and the latter represents the interaction energy when
the two fermions occupy the same site. We choose the nearest-neighbor
hopping $t = 1$ to set our unit of energy. We express all other
energy scales $t_h, U, T$ and $\mu$ in terms of the energy scale $t$.
We fix the hybridization hopping $t_h/t = 0.6$ throughout this work
and studied the properties of our model at half-filling on a bilayer
square lattice with \; $\mathcal{N} = 2 \times L^2$ sites with
periodic boundary conditions. Here $L$ represents the number of sites
in each direction of the square lattice. At half-filling (one fermion
per site), due to particle-hole symmetry in our system, $\mu = 0$.

  In the DQMC simulations, we chose the imaginary-time interval 
$\Delta \tau = 1/20$. Following the steps of the DQMC algorithm
\cite{Blankenbecler1981,Santos2003}, we perform the {Trotter-Suzuki}
decomposition to separate the kinetic and interaction energy
exponentials. For the proposed bilayer band-insulator model, the
kinetic and the interaction exponentials will have the following
expressions:
\begin{equation}
  \label{eqn:YPexpop}
  \left.\begin{aligned}
    \hspace{-3mm} e^{-\Delta\tau \mathcal{K}} = \; && \prod_\sigma
    e^{-\Delta\tau \sum_{\alpha\gamma} \sum_{<{\bf ij}>}
    (c^\dagger_{{\bf i}\alpha\sigma}
    \mathbb{K}^\sigma_{{\bf ij\alpha\gamma}}
     c_{{\bf j}\gamma\sigma} + {\bf h.c.})}     \\
     \hspace{-3mm} e^{-\Delta\tau \mathcal{V}} = \; &&
     e^{\Delta\tau \sum_\alpha \sum_{{\bf i}}
    (- \; U n^\alpha_{{\bf i}\uparrow}
              n^\alpha_{{\bf i}\downarrow}
   + \mu (n_{{\bf i}\alpha\uparrow} +
          n_{{\bf i}\alpha\downarrow}))}
        \end{aligned}
  \right\},
\vspace{-4mm}
\end{equation}
where $(\alpha,\gamma)$ correspond to the layers of the bilayer
system and $c_\alpha (c^\dagger_\alpha)$ is equivalent to the
operators $a \ (a^\dagger)$ and $b \ (b^\dagger)$ for $\alpha=1$ and
$2$, respectively. $\mathbb{K}$ is the kinetic energy matrix whose
elements are given by
\vspace{-3mm}
\begin{equation}
  \label{eqn:YPKijMat}
  \mathbb{K}^\sigma_{{\bf ij\alpha\gamma}} \; = \;
            t_{{\bf ij\alpha\gamma}} \; - \;
           \mu \; \delta_{{\bf ij}} \; \delta_{\alpha\gamma},
\vspace{-4mm}
\end{equation}
with
\begin{widetext}
\begin{equation}
  \label{eqn:YPtnnnij}
  t_{{\bf ij\alpha\gamma}} = 
  \begin{cases}
    \     -t   & \quad \mbox{if ${\bf i}$ and ${\bf j}$
                are the $nn$ for $(\alpha,\gamma) = (1,1)$}  \\
    \     -t'  & \quad \mbox{if ${\bf i}$ and ${\bf j}$
                are the $nnn$ for $(\alpha,\gamma) = (1,1)$} \\
    \quad  t   & \quad \mbox{if ${\bf i}$ and ${\bf j}$
                are the $nn$ for $(\alpha,\gamma) = (2,2)$}  \\
    \quad  t'  & \quad \mbox{if ${\bf i}$ and ${\bf j}$
                are the $nnn$ for $(\alpha,\gamma) = (2,2)$} \\
        -t_h   & \quad \mbox{if ${\bf i}$ and ${\bf j}$
                are the $nn$ for $(\alpha,\gamma) = (1,2)$
                or $(2,1)$}                                  \\
    \quad  0   & \quad \mbox{otherwise}.
  \end{cases}
\end{equation}
\end{widetext}
where $nn$ represents the nearest neighbors, while $nnn$ represents
the next-nearest neighbors. After applying the Hubbard-Stratonovich
transformation for the bilayer band-insulator model, the elements of
the matrix $\mathbb{V}$ and the chemical potential in the kinetic
energy term are modified to
\vspace{-5mm}
\begin{equation}
  \label{eqn:YPexpop}
  \begin{aligned}
    \mathbb{V}^\sigma_{{\bf ij}\alpha\gamma} \; = \; &&
    - \frac{\lambda s_{{\bf i}}}{\Delta\tau} \delta_{{\bf ij}}
     \; \delta_{\alpha \gamma}, \\
     \tilde{\mu} \; = \; &&
     \mu + \frac{\abs{U}}{2}.
  \end{aligned}
\vspace{-3mm}
\end{equation}
Hence, at half filling, we have $\tilde{\mu} = 0$. We calculate
various single-particle quantities such as the momentum distribution,
kinetic energy, and double occupancy, two-particle correlations such
as pair-pair correlations and density-density correlations using the
DQMC simulation. For each data point at the given value of the
inverse temperature $\beta$, the interaction $\abs{U}$, and the
lattice size $L = 10-16$, the simulations were carried out with
$1000-10000$ warm-up sweeps and $10000-50000$ measurement sweeps of
the space-time lattice. These measurement sweeps were divided into
$20$ bins and thus the statistical average over these $20$ bins has
been reported here.

\vspace{-4mm}
\section{Single Particle Properties}
\label{sec:YPSP}
\vspace{-3mm}

  The mean-field analysis and the Gaussian fluctuation theory suggest
that the bilayer band-insulator model undergoes a second-order phase
transition at some critical value of the attractive interaction
$\abs{U_c}$ \cite{Prasad2014}. The fermions start to form pairs as we
tune the on-site attractive interaction through Feshbach resonance.
In the strong-coupling limit $(\abs{U} >> t)$, these pairs get
tightly bound, forming a molecule (boson). Thus, by tuning the
interaction, we go from a band-insulating state to a loosely bound
state of Cooper pairs and then to a tightly bound molecule-forming
state, implying that there is a smooth $BCS-BEC$ crossover extending
from a small to a large value of the interactions. Hence, the ground
state of the system evolves continuously from a $BCS$ state (where
fermions with opposite spins form loose pairs of plane waves with
opposite momenta) to a $BEC$ state of bosonic molecules (where
fermions with opposite spin form tightly bound pairs) when $\abs{U}$
is increased beyond $\abs{U_c}$.

\vspace{-4mm}
\begin{figure}[h]
  \centerline{
  \includegraphics[width=0.96\columnwidth]{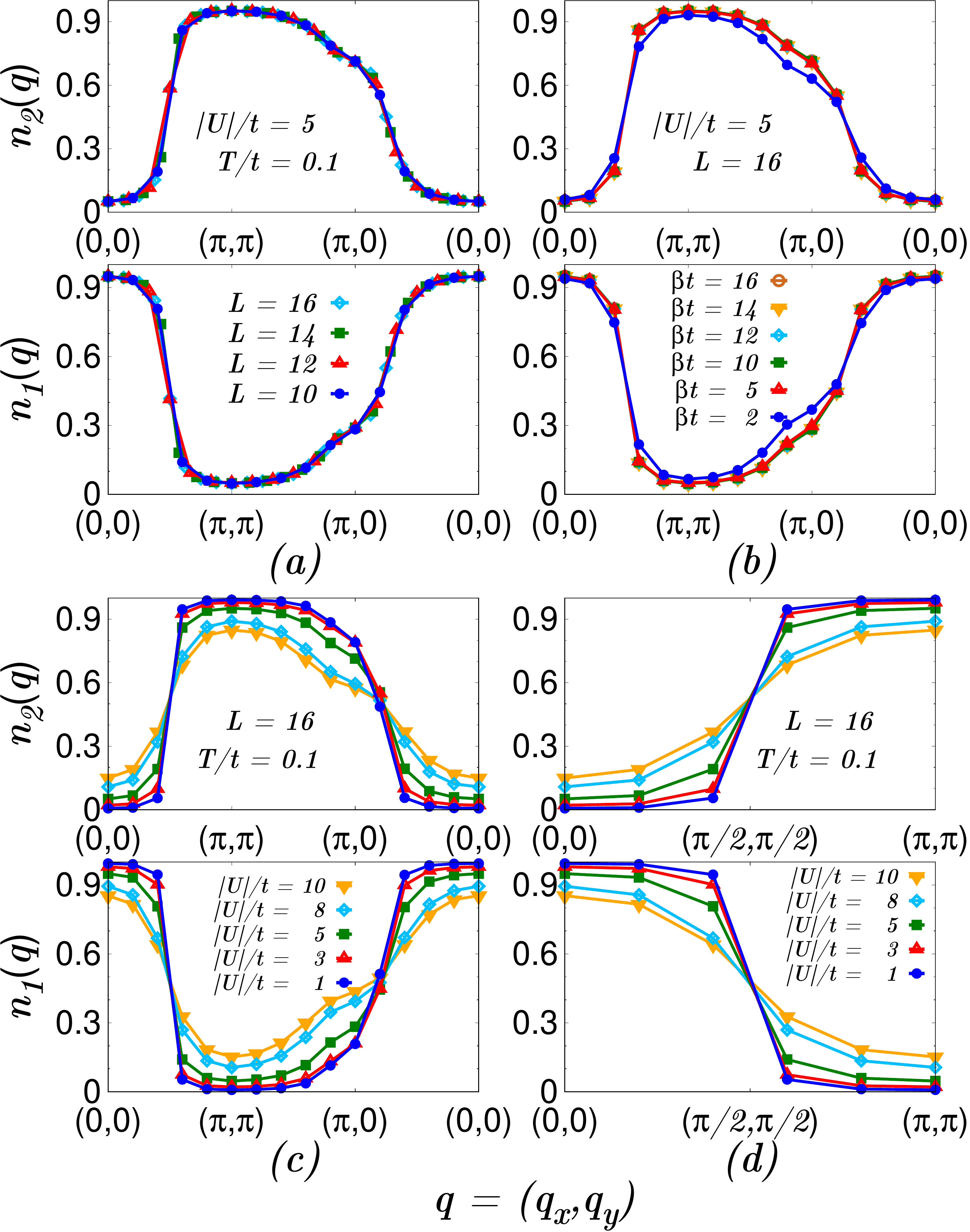}}
  \caption{Momentum distributions. (a) The momentum distributions
    are shown for the interaction strengths $\abs{U}/t=5$ and the
    inverse temperature $\beta t=10$ for various system sizes.
    $n_1({\bf q})$ and $n_2({\bf q})$ are the momentum distributions
    corresponding to layer $A$ and $B$, respectively. It shows weak
    lattice size dependence and the resolution increases with $L$.
    (b) The $n({\bf q})$'s converge to their low temperature value
    as $\beta t > W/t$. Here, $\abs{U}/t=5$ and the system size
    $L=16$. (c) The momentum distributions for various attractive
    interactions ranging from $\abs{U}/t=1$ to $8$; (d) the
    enlarged region, cut perpendicular to the Fermi surface at
    $(\pi/2,\pi/2)$, of (c). We see a sharp Fermi surface at weak
    interaction $\abs{U}$ as the momentum cuts across the Fermi
    surface at ${\bf q}=(\pi/2,\pi/2)$ and it broadens out as
    $\abs{U}/t$ increases.}
  \label{fig:nk12}
\vspace{-4mm}
\end{figure}

\subsection{Momentum Distribution}
\label{subsec:YPMom}

  The Green's function is a fundamental quantity in DQMC where it is
used in various updation processes. The momentum distribution can
be obtained directly from the Green's function via Fourier transform
of the equal-time Green’s function
$G_{{\bf ij}\alpha\beta} =
 \langle c_{{\bf i} \alpha \sigma}
 c^\dagger_{{\bf j} \beta \sigma} \rangle$
as
\vspace{-4mm}
\begin{equation}
  \label{eqn:YPnk12}
  n_{\alpha} ({\bf q}) =
     1 - \frac {1}{2N_\alpha} \sum_{{\bf i},{\bf j},\sigma}
     e^{i {\bf q . l^\alpha_{{\bf ij}}}}
     \langle c_{{\bf i} \alpha \sigma}
     c^\dagger_{{\bf j} \alpha \sigma} \rangle
\vspace{-2mm}
\end{equation}
where ${\bf l^\alpha_{{\bf ij}}} = {\bf l(i,\alpha) - l(j,\alpha)}$
and $N_\alpha (= L^2)$ represents the number of sites in the
$\alpha$th layer.

\begin{figure}[h]
  \centerline{
  \includegraphics[width=0.98\columnwidth]{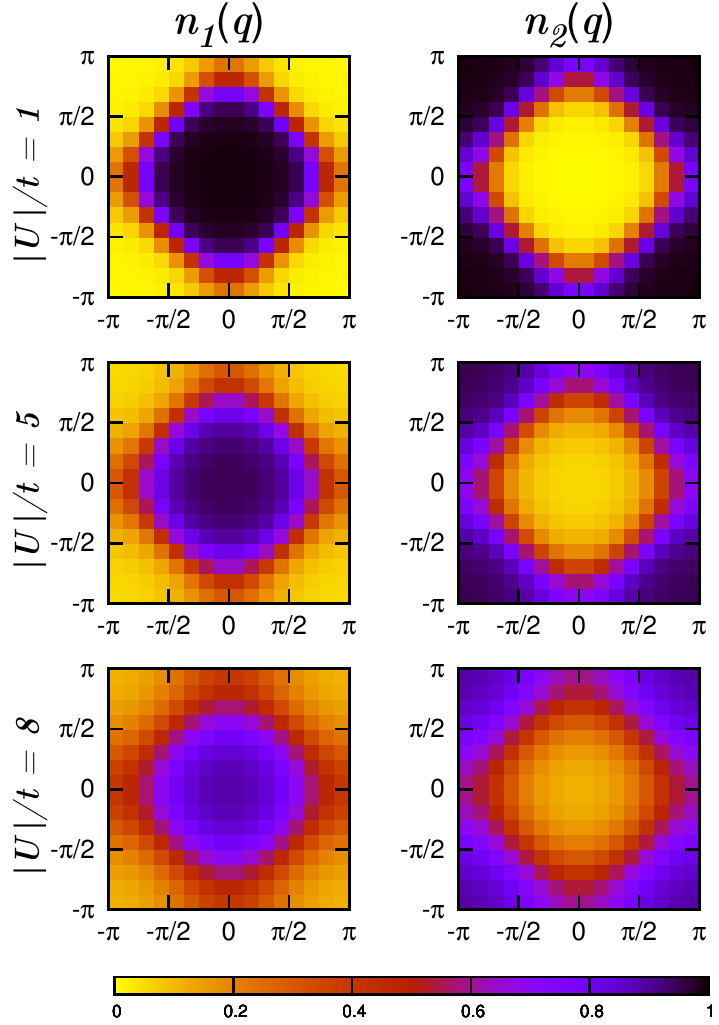}}
  \caption{Color contour plot depiction of the momentum
    distributions. $n_1({\bf q})$ and $n_2({\bf q})$ at half filling
    for $\abs{U}/t = 1, 5$, and $8$. The lattice size $=2 \times 16
    \times 16$ and the inverse temperature $\beta t=10$.}
  \label{fig:nk12pmd}
\end{figure}

  \Figg{fig:nk12} shows the momentum distribution of our bilayer
square lattice around the irreducible part of the Brillouin zone (BZ)
for various system sizes $\bigl [$\Fig{fig:nk12}(a)$\bigr ]$, at
different temperatures $\bigl [$\Fig{fig:nk12}(b)$\bigr ]$, and for
different interactions $\abs{U}$ $\bigl [$\Figs{fig:nk12}(c) and
\Figm{fig:nk12}(d)$\bigr ]$. At $\abs{U} = 0$ and at half filling,
$n_1({\bf q}) \ [n_2({\bf q})] = 1 (0)$ inside and
$n_1({\bf q}) \ [n_2({\bf q})] = 0 (1)]$ outside a square with
vertices $(\pi,0), (0,\pi), (-\pi,0)$, and $(0,-\pi)$ within the BZ.
\Figg{fig:nk12}(a) shows that the momentum distribution has a weak
lattice size dependence and its resolution increases with $L$. In
\Fig{fig:nk12}(b), we show that the $n({\bf q})$'s converge to their
respective low-temperature values as $\beta t > W/t$, where
$W(=8.8 \; t)$ is the bandwidth of the noninteracting energy
dispersion. We find that the smearing due to the finite-temperature
effects is very small above $\beta \; t=8.8$. In \Fig{fig:nk12}(c),
we see a sharp Fermi surface at weak interaction $\abs{U}$ as the
momentum cuts across the Fermi surface at ${\bf q}=(\pi/2,\pi/2)$,
whose enlarged plot is shown in \Fig{fig:nk12}(d), which focuses on
the region near the Fermi surface point $(\pi/2,\pi/2)$. It shows
that the distribution broadens out as the interaction $\abs{U}$
increases. In \Fig{fig:nk12pmd}, we show a sequence of color contour
plots for the lattice size $L=16$ at $\abs{U}/t=1, \; 5$, and $8$.

\vspace{-4mm}
\subsection{Pair Formation}
\label{subsec:YPPF}
\vspace{-4mm}

  The existence of the molecule formation along the $BCS-BEC$
crossover with the increase in the interaction strength $\abs{U}$
comes from the evolution of the double occupancy (density of
on-site pairs), which can also be measured in experiments with
ultracold fermions \cite{Jordens2008}. The double occupancy $D$
can be defined as
\vspace{-2mm}
\begin{equation}
  \label{eqn:Dble}
  D =
    \frac {1}{N} \sum_{{\bf i}\alpha}
    \mean{n_{{\bf i}\uparrow}^\alpha
    n_{{\bf i}\downarrow}^\alpha}
\vspace{-2mm}
\end{equation}
where the summation $i$ runs over the number of sites for layer
$\alpha$ for both layers $\alpha = 1$ and $2$, and $N (= 2L^2)$
represents the total number of sites.

In the noninteracting limit $\bigl (\abs{U}/t = 0 \bigr )$, both
spin-up and spin-down particles are uncorrelated so
$\sum_{{\bf i}\alpha} \mean{n_{{\bf i}\uparrow}^\alpha
n_{{\bf i}\downarrow}^\alpha}/N = \sum_{{\bf i}\alpha}
\mean{n_{{\bf i}\uparrow}^\alpha}
\mean{n_{{\bf i}\downarrow}^\alpha}/N = 1/4$ at half filling. As we
increase the attractive interaction by tuning the scattering rate,
the spin-up and spin-down particles become correlated. In the
strong-coupling limit $\bigl (\abs{U}/t \rightarrow \infty \bigr )$,
the large attractive interaction locks the fermions in bound on-site
pairs and hence they form a charge density wave where the fermion
pairs occupy alternate sites of the bilayer lattice at half filling
to minimize the energy by a virtual hopping process of the order of
$t^2/\abs{U}$. In this limit, at half filling, the on-site pair
density or the double occupancy will be equal to the number of up- or
down-spin fermions which is $0.5$ at half filling: since $n=1$,
$\sum_{{\bf i}\alpha} \mean{n_{{\bf i}\uparrow}^\alpha
n_{{\bf i}\downarrow}^\alpha}/N = \sum_{{\bf i}\alpha}
\mean{n_{{\bf i}\uparrow}^\alpha}/N = \sum_{{\bf i}\alpha}
\mean{n_{{\bf i}\downarrow}^\alpha}/N = 1/2$.

\begin{figure}[t]
  \centerline{
  \includegraphics[width=0.98\columnwidth]{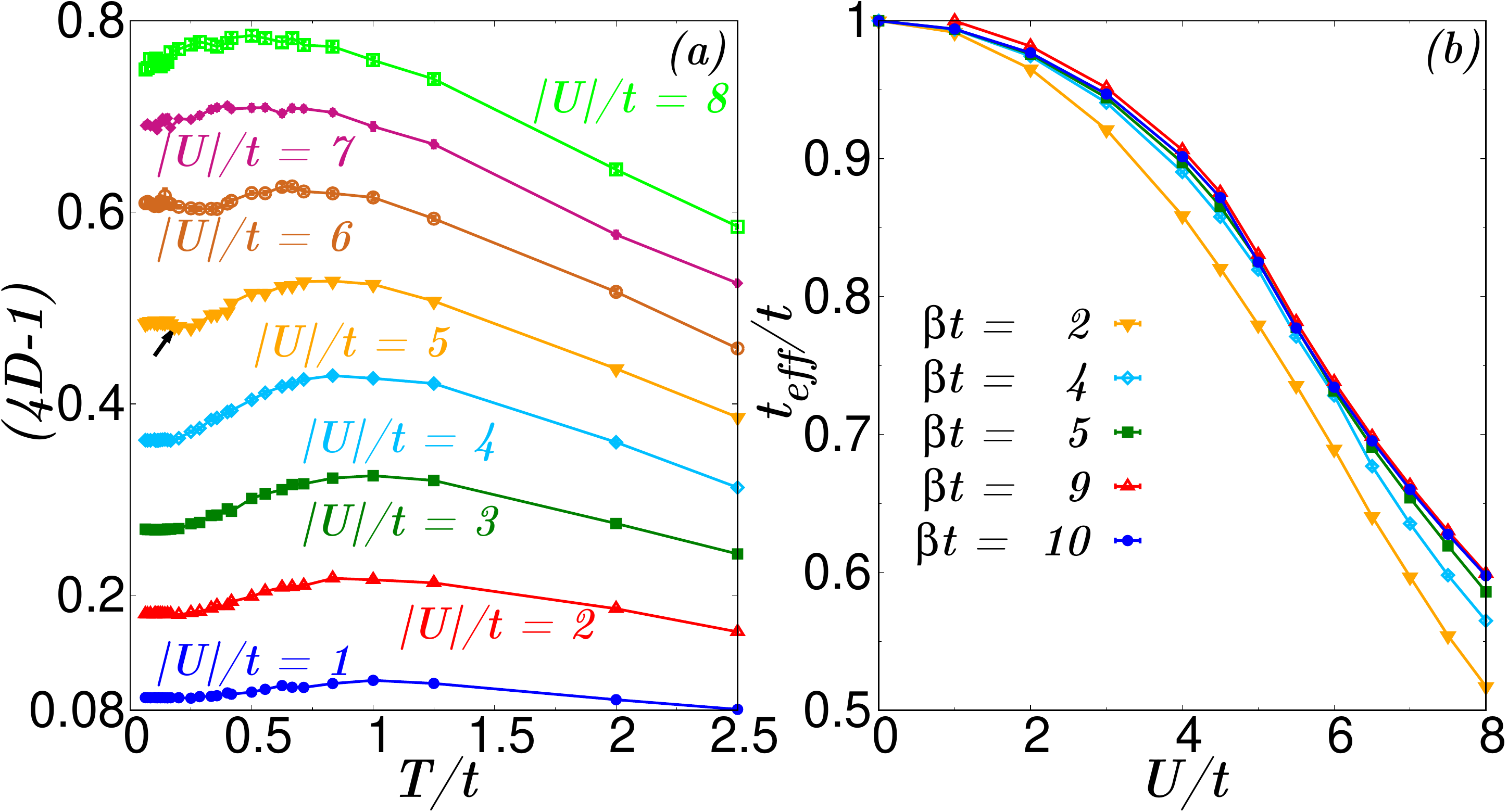}}
  \caption{The evolution of the rescaled double occupancy with the
    temperature $T/t$ for various interaction strengths $\abs{U}/t$.
    The system size is $L=16$. The black arrow marks the transition
    temperature $T_c/t=0.17$ at $\abs{U}/t=5$, estimated from the
    finite-size scaling analysis. The rescaled double occupancy
    reaches up to $0.8$ at $\abs{U}/t=8$, which is an indication of
    the molecule (tightly bound pairs) formation. A local maxima has
    been observed as we go from the high-temperature to the
    low-temperature regime at the intermediate-temperature scale
    $T/t \sim 1$. (b) Effective hopping: as the interaction energy
    $\abs{U}$ increases, the effective hopping declines. The rate of
    decrease of the effective hopping with the increase in the
    interaction strength is the same for all $\beta t > W/t (8.8/t)$.
    Here we show $t_{eff}/t$ for $L=16$ for different values of
    inverse temperature $\beta t$.}
  \label{fig:YPdbleoccKEeff}
\end{figure}

  In \Fig{fig:YPdbleoccKEeff}(a), we have plotted the
rescaled double occupancy,
\vspace{-2mm}
\begin{equation}
  \label{eqn:RescDble}
  \tilde{D} =
  \frac {D - \mean{n_{{\bf i}\uparrow}^\alpha}^2}
  {\mean{n_{{\bf i}\uparrow}^\alpha} -
   \mean{n_{{\bf i}\uparrow}^\alpha}^2} = (4 \; D - 1)
\vspace{-2mm}
\end{equation}
as a function of temperature $T$ for various interaction strengths.
As the attractive interaction increases from zero to infinity, the
rescaled double occupancy increases from $0$ to $1$. In the
$\abs{U} \rightarrow \infty$ limit, all the fermions get paired up
and hence $\tilde{D}$ approaches the value $1$. The rescaled double
occupancy reaches upto $0.8$ at $\abs{U}/t = 8$, which is an
indication of the formation of bosonic molecules (tightly bound pairs
of fermions of opposite spins). In the high-temperature limit
$\bigl (T/t >> 1 \bigr )$, independent of the interaction strength,
the double occupancy approaches the noninteracting values, i.e.,
$0.25$ in the half-filled case. As we decrease the temperature, the
fermions start to pair up due to the increasing effect of the
attractive interaction. Hence the double occupancy increases with the
decrease in temperature $T$. We observe a local maxima as we go from
the high-temperature to the low-temperature regime, implying the
increase in the double occupancy with the decrease in the
temperature. We see the local maxima at the intermediate temperature
scale $\bigl (T/t \sim 1\bigr )$, where the kinetic energy competes
with the on-site interaction and destabilizes the double occupancy in
the weak- and intermediate-coupling regimes. So the double occupancy
decreases a little and saturates in the weak-coupling regime, after
reaching its ground-state value as $T/t \rightarrow 0$. But after a
certain critical interaction strength $\abs{U_c}$, we observe that as
we decrease the temperature further, the double occupancy increases
again after a critical temperature $T_c$ and then saturates to its
low-temperature value. This indicates that the system goes into a
superfluid phase from the band-insulating phase at this critical
interaction strength and temperature. We estimate these critical
values through the scaling analysis discussed at the end of this
paper.

\vspace{-6mm}
\subsection{Kinetic Energy}
\label{subsec:YPKE}
\vspace{-4mm}

\begin{figure}[h]
  \centerline{
  \includegraphics[width=0.98\columnwidth]{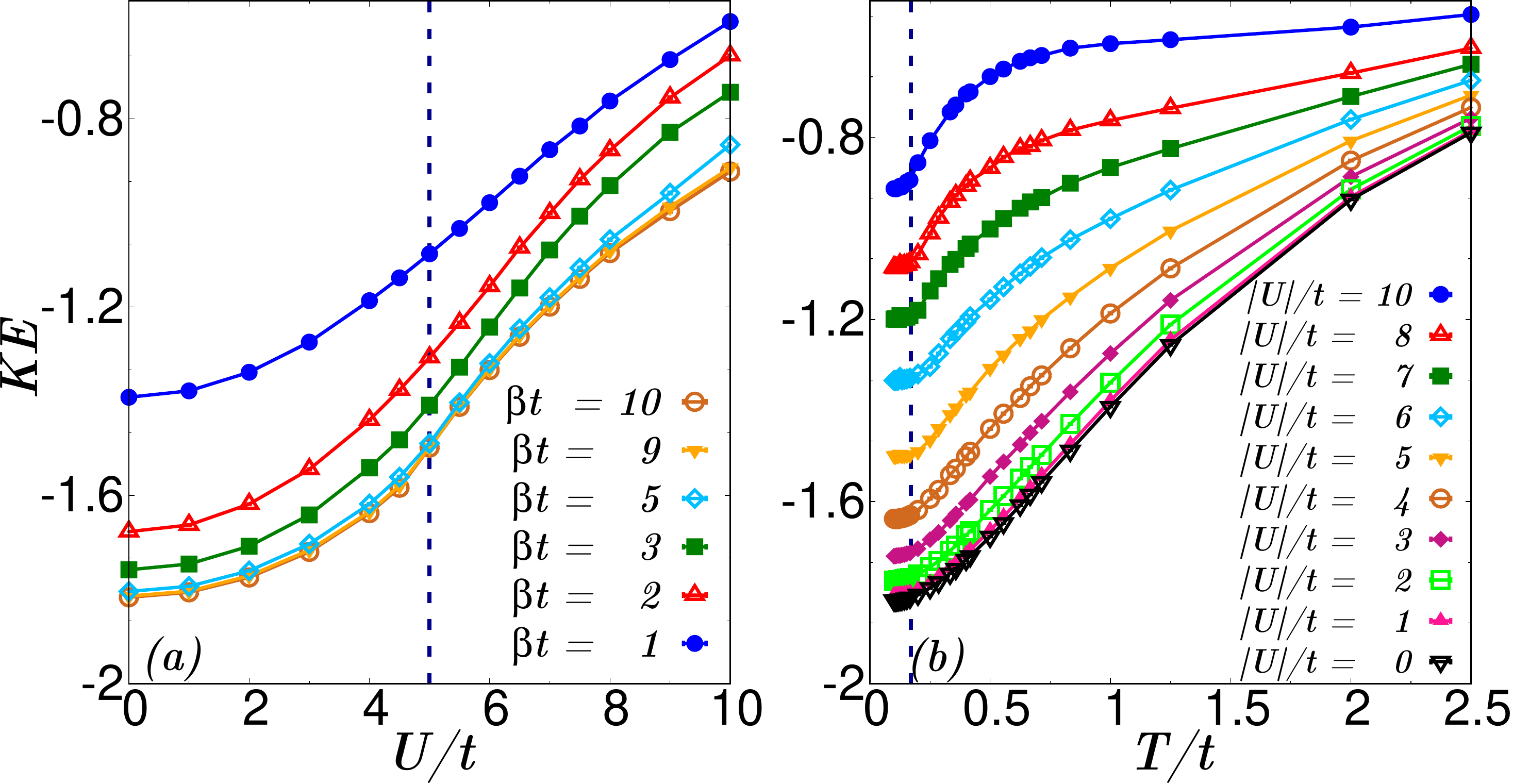}}
  \caption{The evolution of the kinetic energy as a function of $(a)$
    interaction strength $\abs{U}/t$ for various temperatures and
    $(b)$ temperature $T/t$ for different interaction strengths. The
    kinetic energy (KE) increases continuously with the increase in
    the attractive interaction, approaching its zero-temperature
    value for $\beta t > W/t (8.8/t)$. The evolution of KE with
    temperature $T$ at weak couplings shows a similar behavior as
    the free-fermion case, while at strong couplings, we observe a
    sharp fall of the KE at low temperatures before it approaches its
    zero-temperature value. Here the system size is $L=16$. In (a),
    the vertical dashed line show the value of $\abs{U_c}/t=5$ in the
    ground-state, and in (b), it corresponds to $T_c/t=0.17$ at
    $\abs{U}/t=5$.}
  \label{fig:YPKE}
\end{figure}

  Another single-particle quantity of interest is the effective
hopping, defined as
\vspace{-3mm}
\begin{equation}
  \label{eqn:teff}
  \frac {t_{eff}}{t} =
  \frac {\langle \mathcal{H}_K \rangle_U}
  {\langle \mathcal{H}_K \rangle_{U = 0}},
\vspace{-3.5mm}
\end{equation}
which measures the ratio of the kinetic energy at finite $\abs{U}$ to
its value when there is no on-site interaction, i.e., $U=0$. In
\Fig{fig:YPdbleoccKEeff}(b), we show the effective hopping as a
function of the interaction strength $\abs{U}$ for $L=16$ at various
temperatures. As the on-site attractive interaction $\abs{U}$
increases, the effective hopping decreases. This is due to the fact
that the kinetic energy increases from larger negative value to
the lower negative value with the increase in the interaction,
as shown in \Fig{fig:YPKE}(a). The rate of decrease of the effective
hopping with the increase in the interaction strength is the same for
all $\beta t > W/t (8.8/t)$, implying that the kinetic energy has
reached its low-temperature value, as seen in \Fig{fig:YPKE}(a). In
\Fig{fig:YPKE}(b), we show the evolution of the kinetic energy as a
function of temperature $T/t$ for different interaction strengths for
a system of $512$ sites. We see that the evolution of the kinetic
energy with temperature $T$ at weak couplings shows a similar
behavior as the free-fermion case, while for intermediate couplings,
the rate of decrease in kinetic energy with the decrease in
temperature is slow as compared to the weak-coupling behavior. In the
strong-coupling limit, at low temperatures, we see a sharp drop in
the kinetic energy which then approaches its zero-temperature value,
which is due to the reason that the kinetic energy at strong
couplings is determined by the effective hopping of the paired
fermions which form a bound state in this limit.

%----------------------------------------------------------------------------------------
%       SECTION 3
%----------------------------------------------------------------------------------------

\vspace{-6mm}
\section{Two Particle Correlations}
\label{sec:YPTwoParticleCorr}
\vspace{-4mm}

  We now turn to the two-particle properties focusing on the pair
correlations in the bilayer band insulator model. We will also
discuss the density-density correlations to see the possibility of
formation of the charge density wave (CDW) state at a large value of
the on-site attractive interaction. We also estimate the critical
strength $\abs{U_c}$ and the critical temperature $T_c$ through the
finite-size scaling analysis.

\vspace{-5mm}
\subsection{Pair-Pair Correlations}
\label{subsec:YPPairCorr}
\vspace{-4mm}

  We know that the long-range order (or a quasi-long-range order
for a superfluid at finite temperature) in the pair-pair
correlation function in the Bose-Einstein condensate state
signifies a phase coherence between pairs. To study this
behavior, we define the {\it equal-time} $s$-wave pairing
$P_s^{\alpha \gamma} ({\bf i}, {\bf j})$ for the bilayer model as
\vspace{-1.5mm}
\begin{equation}
  \label{eqn:Pswave}
  P_s^{\alpha \gamma} ({\bf i},{\bf j}) =
  \mean{\Delta_s ({\bf i},\alpha)
  \Delta^\dagger_s ({\bf j},\gamma) + {\bf h.c.}}
\end{equation}
where the local {\it pair-field} operator
$\Delta_s ({\bf i},\alpha)$, defined as
\vspace{-1.5mm}
\begin{equation}
  \label{eqn:YPDeltaPair}
  \Delta_s ({\bf i},\alpha) =
    c_{{\bf i},\alpha \downarrow} c_{{\bf i},\alpha \uparrow},
\end{equation}
annihilates a {\it pair} of fermions on site ${\bf i}$ of layer
$\alpha$ of the bilayer-square lattice. We also define the
associated pair structure factor as
\vspace{-1.5mm}
\begin{equation}
  \label{eqn:PSF}
  S_s ({\bf q}) =
    \frac{1}{N} \sum_{\alpha \gamma} \sum_{{\bf ij}}
    e^{{\it i} {\bf q}.{\bf l^{\alpha \gamma}_{{\bf ij}}}} \
    P_s^{\alpha \gamma} ({\bf i,j})
\vspace{-1.5mm}
\end{equation}
where ${\bf l^{\alpha\gamma}_{{\bf ij}}} =
{\bf l(i,\alpha) - l(j,\gamma)}$.

  The pair structure factor diverges linearly with the system size
$N$ when the long-range order is achieved. As the bilayer
band-insulator system undergoes a finite-temperature
Berezinski-Kosterlitz-Thouless (BKT) transition into a
superfluid phase \cite{Prasad2014}, hence for $0 < T \leq T_c$,
we expect that
\vspace{-1.5mm}
\begin{equation}
  \label{eqn:PPCScal}
  P_s ({\bf l}) \; \sim \; {\bf l}^{-\eta(T)},
\vspace{-1.5mm}
\end{equation}
with the separation ${\bf l}=\abs{{\bf i}-{\bf j}}$, where
${\bf i,j}$ refers to sites either from layer $A$ or $B$. The
critical exponent $\eta(T)$ for a $BKT$ transition in a homogeneous
system is known to increase monotonically with temperature between
$\eta(0) = 0$ and $\eta(T_c) = 1/4$ \cite{Berezinskii1972,
Kosterlitz1973}.

\begin{figure}[h]
  \centerline{
  \includegraphics[width=0.96\columnwidth]{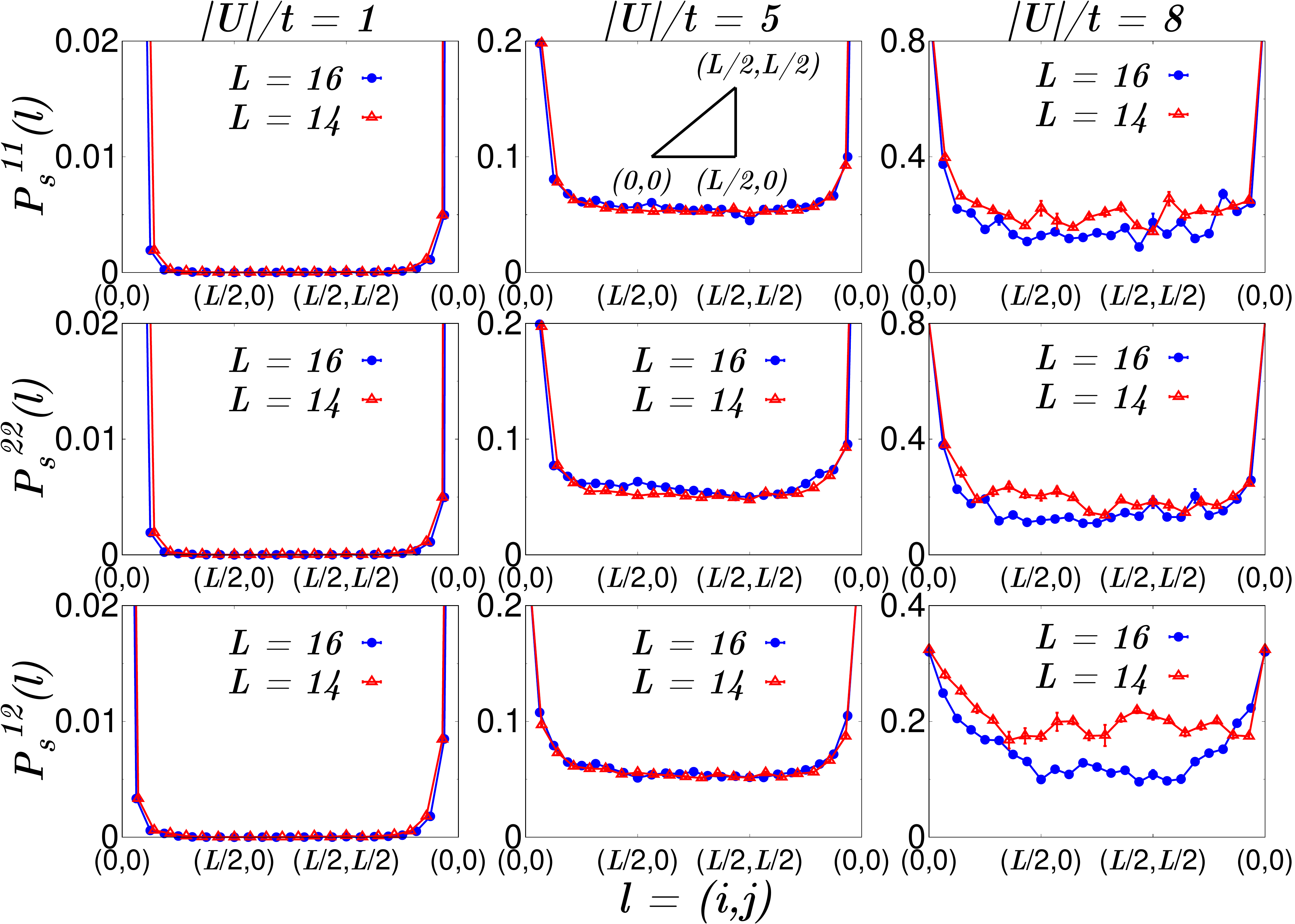}}
  \caption{The dependence of the ground-state pair-pair correlation
    function on the separation ${\bf l}$ for two different lattice
    sizes $L=14$ and $L=16$, in a bilayer band-insulator model. The
    separation ${\bf l}$ follows a trajectory along the $x$ axis to
    maximal $x$ separation $(\frac{L}{2},0)$ on a lattice with
    periodic boundary conditions, and then to $(\frac{L}{2},
    \frac{L}{2})$ before returning to separation $(0,0)$. The
    correlation functions converge to a nonzero value at large
    separations at $\abs{U}/t=5$ and $8$, providing clear evidence
    for the long-range order. We see that the finite-size effects are
    modest.}
  \label{fig:YPpairlxly}
\end{figure}

  Thus we can obtain the finite-size scaling behavior of the $s$-wave
pair structure factor upon integrating $P_s ({\bf l})$ over a
two-dimensional system of {\it linear} dimension $L$. Hence,
$S_s (\equiv S_s ({\bf q} = 0))$ will scale as \cite{Moreo1991}
\vspace{-1.5mm}
\begin{equation}
  \label{eqn:PSScal}
  L^{2-\eta(T_c)} \; S_s \; \sim \; f(L/\xi),
  \quad \quad L >>1, \; T \rightarrow T_c^+
\vspace{-1.5mm}
\end{equation}
with $\xi \sim exp[A/{(T-T_c)^{1/2}}]$ the correlation length of the
infinite system where $A$ is of the order of unity. In the
thermodynamic limit, one can recover $S_s \sim \xi^{7/4}$.

  In \Fig{fig:YPpairlxly}, we have shown the dependence of the pair
correlation function on the separation ${\bf l}$ for two different
lattice sizes $L=14$ and $16$. The separation ${\bf l}$ follows a
trajectory along the $x$ axis to maximal $x$ separation
$(\frac{L}{2},0)$ on a lattice with periodic boundary conditions, and
then to $(\frac{L}{2},\frac{L}{2})$ before returning to separation
$(0,0)$. We observe that there is no pair-pair correlation when
$\abs{U}/t=1$ as the system is still in the band-insulating state.
For $\abs{U}/t = 5$ and $8$, there is a finite nonzero pair-pair
correlation, implying the existence of the long-range order at these
interactions. We see that the finite-size effects are modest.

\begin{figure}[b]
  \centerline{
  \includegraphics[width=0.98\columnwidth]{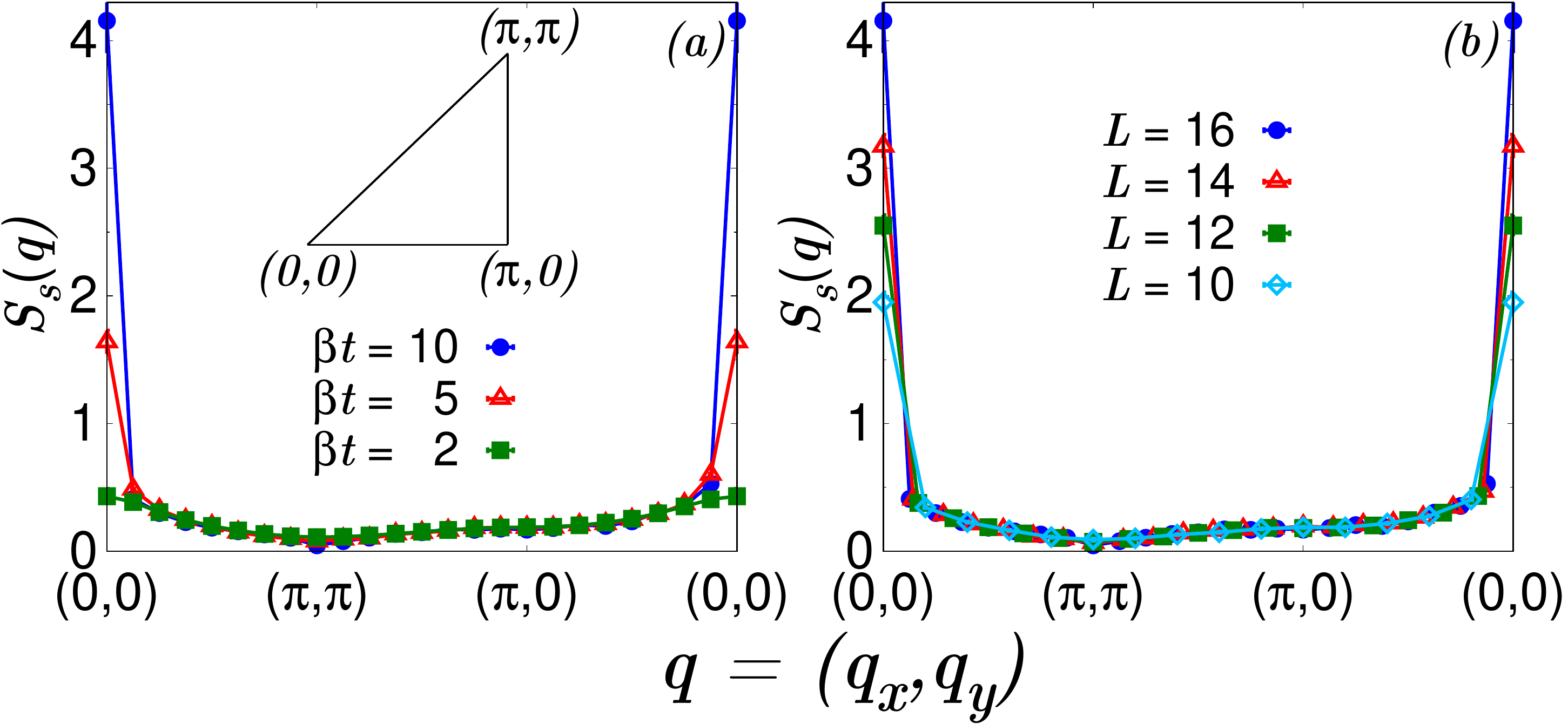}}
  \caption{$s$-wave pair structure factor $S_s ({\bf q},\tau=0)$ at
    $\abs{U}/t = 5$ for (a) various temperatures $(\beta \ t=2-10)$
    on a $2\times16\times16$ lattice and (b) various system sizes
    $(L=10-16)$ at $T/t=0.1$. We see that the ${\bf q}=0$ mode
    becomes more and more singular in both results as we decrease the
    temperature or increase the system size.}
  \label{fig:pairkxky}
\end{figure}

  To have a clearer understanding, in \Fig{fig:pairkxky}, we show the
equal-time $s$-wave pair structure factor $S_s({\bf q},\tau=0)$ for
$\abs{U}/t=5$ [\Fig{fig:pairkxky}(a)] for various temperatures
$(\beta \ t=2-10)$ on a $2\times16\times16$ lattice and
[\Fig{fig:pairkxky}(b)] for various system sizes $(L=10-16)$ at
$T/t=0.1$. We see that the ${\bf q}=0$ mode becomes more and more
singular in both results as we decrease the temperature or increase
the system size, which is a characteristic feature of growing
$s$-wave pairing correlations. The dependence of the ${\bf q}=0$ mode
on the system size indicates that we need to understand the
finite-size scaling behavior of $S_s$ due to the limitations on
lattice size $(L\leq 16)$ imposed by the DQMC algorithm.

\vspace{-6mm}
\subsubsection*{{\bf Finite-size Scaling}}
\label{subsubsec:YPFSSP}
\vspace{-4mm}

\begin{figure}[t]
  \centerline{
  \includegraphics[width=0.98\columnwidth]{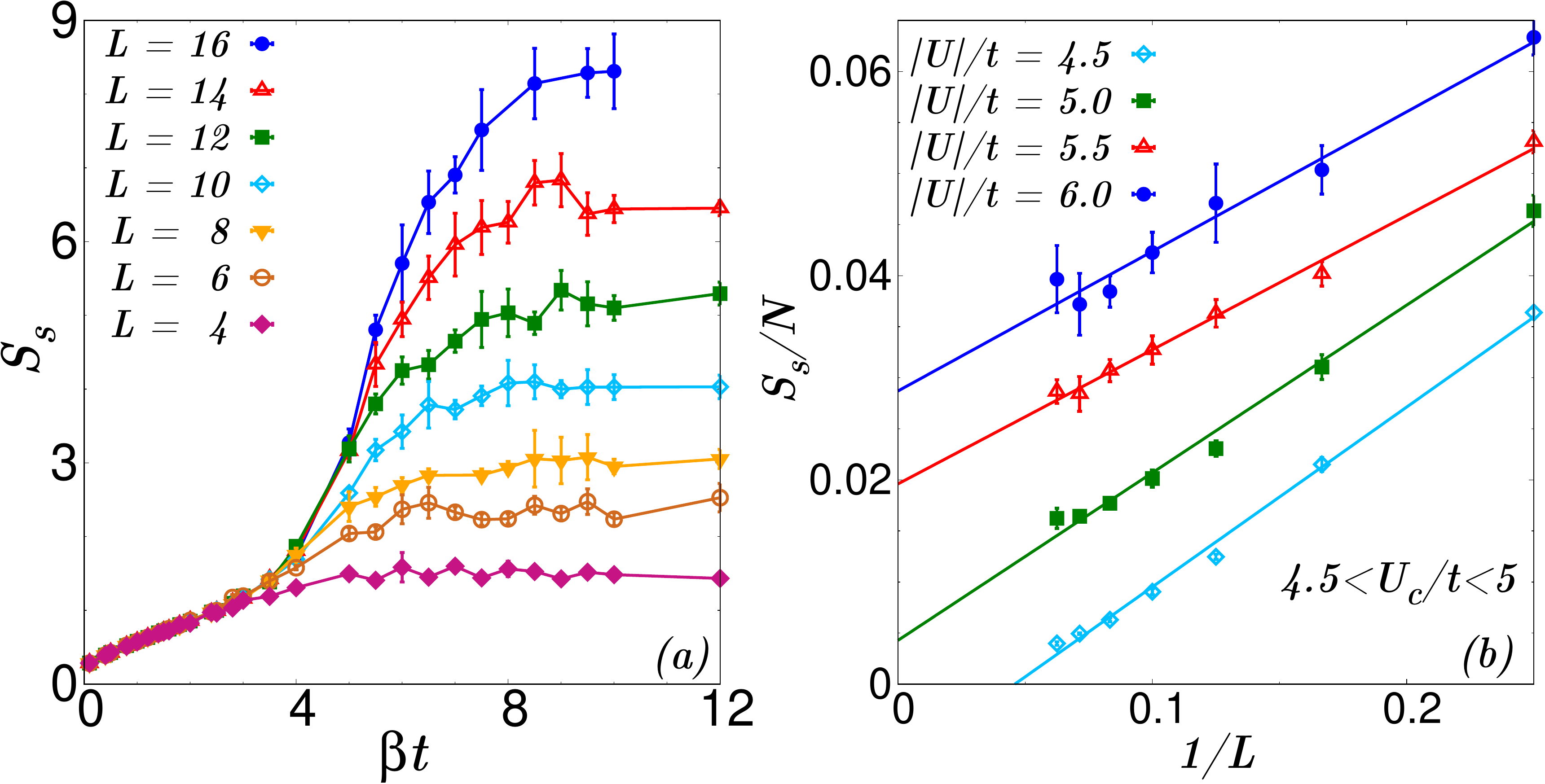}}
  \caption{(a) The evolution of the $s$-wave pair structure factor
    $S_s$ with the inverse temperature $\beta$ for different system
    sizes at $\abs{U}/t=5$. $S_s$ increases at low temperatures,
    saturating at a value which increases with the size of the
    lattice. (b) Ground state: Finite-size scaling of the $s$-wave
    pair structure factor for various interaction strengths. The
    symbols are the DQMC results and the dashed lines are the
    extrapolation performed via a linear least-squares fit for each
    $\abs{U}$. We observe that $\abs{U_c}/t \sim 5$.}
  \label{fig:YPPsvsT_L}
\end{figure}

  To understand the finite-size scaling behavior of $S_s$, we obtain
the low-temperature limit of the pair structure factor by decreasing
the temperature until we observe a plateau which signals that we have
reached the $T=0$ value of the pair structure factor.
\Figg{fig:YPPsvsT_L}(a) shows the evolution of $S_s$ with the inverse
temperature $\beta \ t$ for $\abs{U}/t=5$ and for different lattice
sizes. Here we see that $S_s$ increases at low temperatures and
saturates to a value which increases with the size of the lattice.
For $\beta \leq 4$, pair correlations are short range, so the pair
structure factor is independent of the lattice size. As we decrease
the temperature, the point at which the pair structure factor begins
to grow with the lattice size indicates the temperature at which the
correlation length $\xi$ becomes large as compared to the lattice
size $L$.

  {\it Ground State: $\abs{U_c}$.} The important characteristic of
the superfluid state in our bilayer model is that the system displays
long-range order in the ground state, and hence Huse's argument
\cite{Huse1988} of the ``spin-wave scaling'' is expected to hold
\cite{Paiva2004},
\vspace{-2mm}
\begin{equation}
  \label{eqn:PsDelta}
  \frac{S_s}{N} = \varDelta_0^2 + \frac{C (U)}{L}
\vspace{-2mm}
\end{equation}
where $\varDelta_0$ is the superfluid order parameter at zero
temperature and $C$ is a constant which depends on the interaction
strength $\abs{U}$.

  The superfluid order parameter $\varDelta_0$ can also be extracted
from the {\it equal-time} $s$-wave pair-pair correlation function
$P_s ({\bf l})$ for the two most distant points on a lattice, i.e.,
having ${\bf R} = (L/2,L/2)$ \cite{Scalettar1999}, with a similar
spin-wave theory correction,
\vspace{-2mm}
\begin{equation}
  \label{eqn:PairDelta}
  P_s ({\bf R}) = \varDelta_0^2 + B(U) L.
\vspace{-2mm}
\end{equation}
We expect that $B<C$ since the structure factor includes the pair
correlations at short distances which markedly exceed
$\varDelta_0^2$, in addition to the finite lattice effects at larger
length scales \cite{Varney2009}.

  In \Fig{fig:YPPsvsT_L}(b), we perform the finite-size
scaling of the $s$-wave pair structure factor $S_s$ for various
interaction strengths. We can approximate the superfluid order
parameter by the intercept along the $y$ axis as evident from
\Eqn{eqn:PsDelta}. Thus we observe that the zero-temperature order
parameter is nonzero for the interaction $\abs{U}/t \geq 5$. Hence
the critical interaction strength $\abs{U_c}/t \sim 5$.

\begin{figure}[t]
  \centerline{
  \includegraphics[width=0.96\columnwidth]{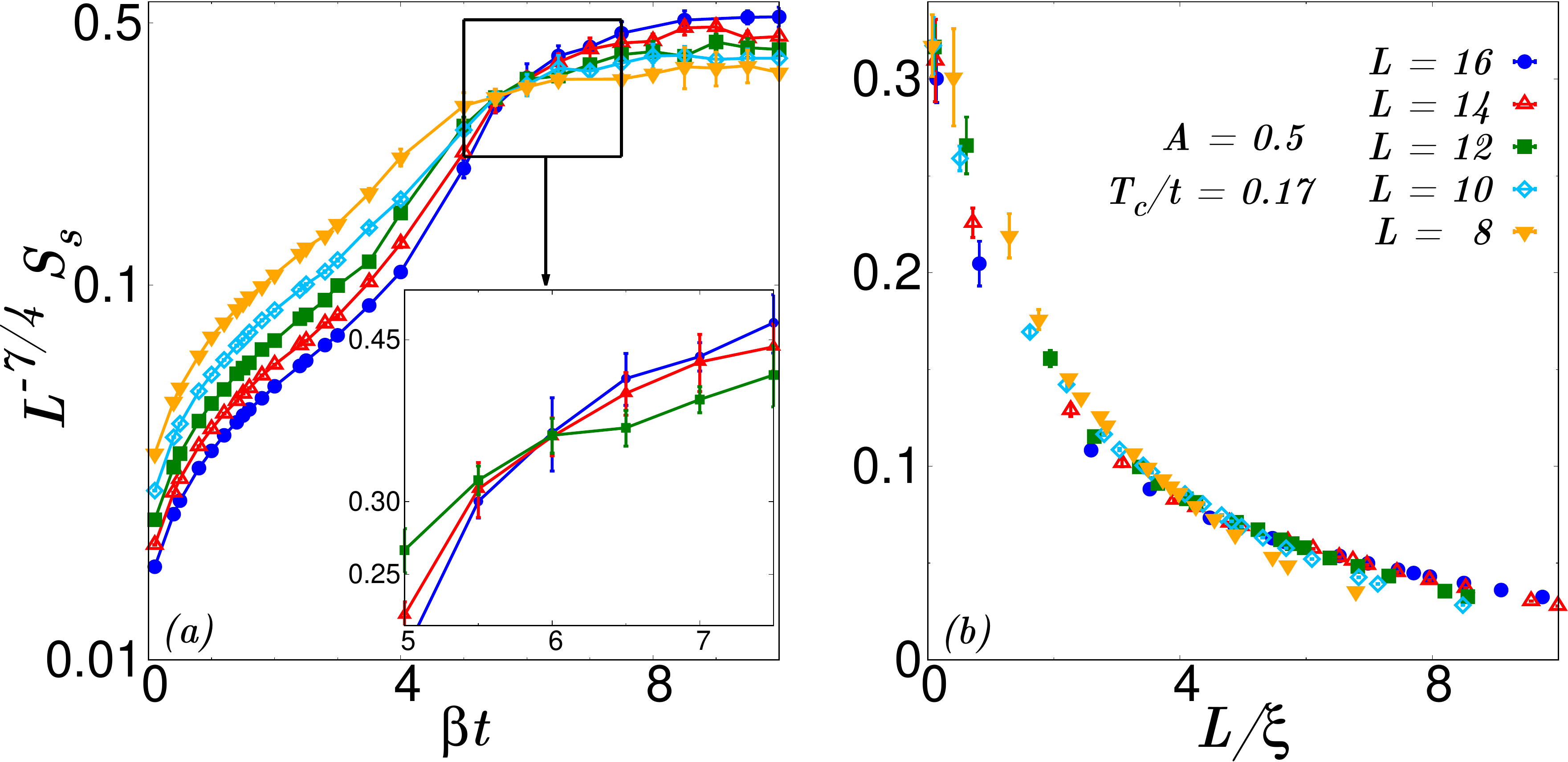}}
  \caption{(a) Rescaled $S_s$ as a function of inverse temperature
    $\beta$ at $\abs{U}/t = 5$ for different system sizes. The inset
    shows the enlarged region around $\beta \ t=5-7.5$. All the
    curves for different system sizes intercept each other at
    $\beta \ t\sim 6$. (b) Rescaled $S_s$ plotted against $L/\xi$ at
    $\abs{U}/t=5$ for different system sizes. All the curves for
    different system sizes collapse to a single curve at $A=0.5$ and
    $T_c/t=0.17$.}
  \label{fig:YPScaledPs}
\end{figure}

\begin{figure}[b]
  \centerline{
  \includegraphics[width=0.96\columnwidth]{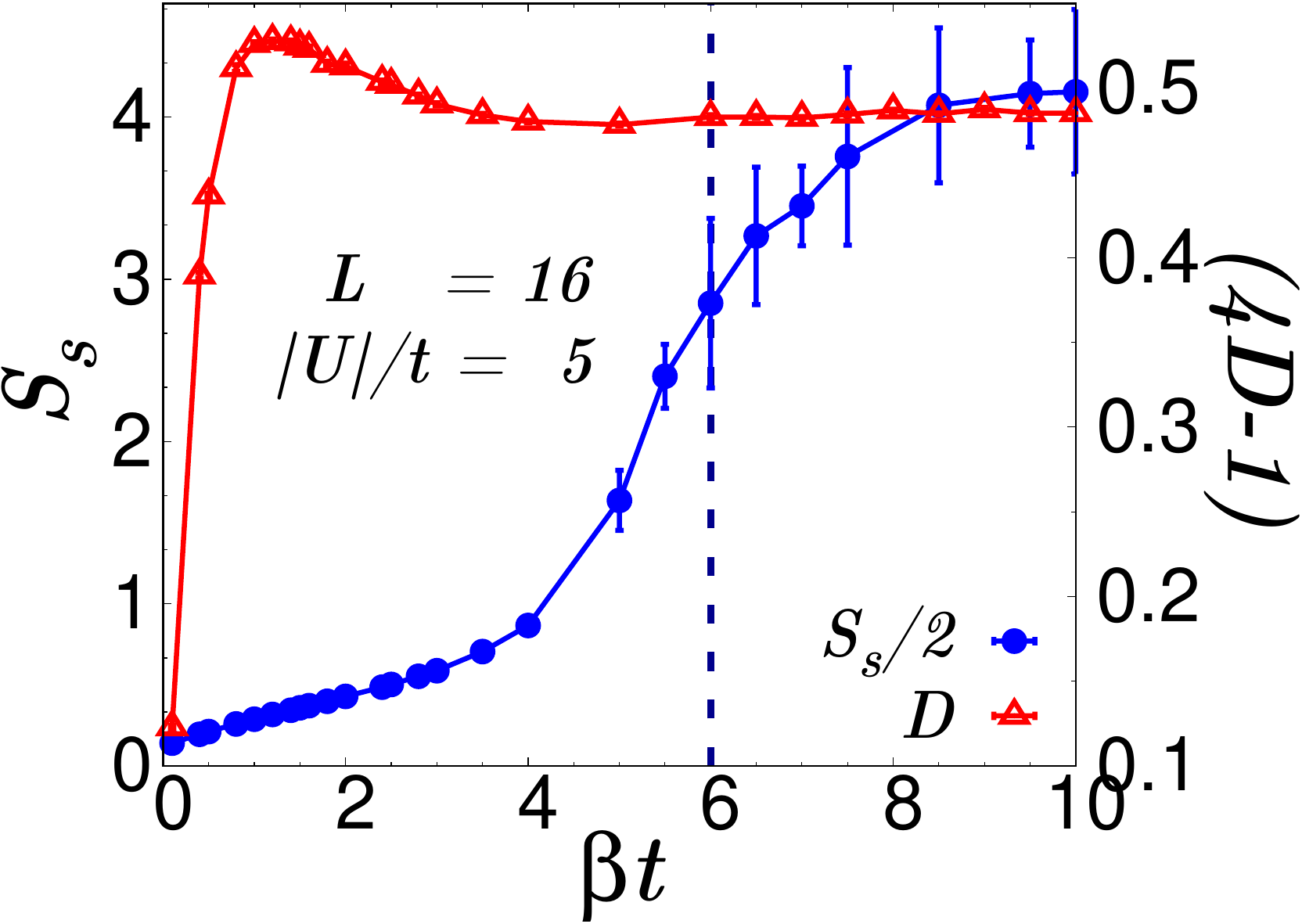}}
  \caption{The evolution of the $s$-wave pair structure factor and
    the rescaled double occupancy with inverse temperature $\beta$ at
    $\abs{U}/t=5$ and $L=16$. Two different energy scales are clearly
    identified. $S_s$, signaling the emergence of the phase
    coherence, saturates at $\beta \ t \sim 8$, whereas $\tilde{D}$,
    signaling the molecule formation, saturates at
    $\beta \ t \sim 3$.}
  \label{fig:YPdblePsvsU}
\end{figure}

  {\it Estimation of $T_c$.} We can extract $T_c$ from the pair
structure factor $S_s$ through a ``phenomenological renormalization
group" analysis \cite{Nightingale1982,Santos1981}. As we know,
$\xi=\infty$ at $T_c$ and $\rightarrow \infty$ for all $0<T<T_c$ [see
\Eqn{eqn:PSScal}]. Thus, at $T_c$, $L^{-7/4} \; S_s (L,\beta_c)$
becomes a constant, independent of the system size. Hence all the
curves for different system sizes, in the plot of the rescaled pair
structure factor $L^{-7/4} \; S_s (L,\beta)$, should {\it intersect}
at $\beta=\beta_c$, when plotted as a function of $\beta$. In
\Fig{fig:YPScaledPs}(a), we show the rescaled pair structure factor
$L^{-7/4} \; S_s (L,\beta)$ as a function of the inverse temperature
$\beta$ for $\abs{U}/t=5$ for different lattice sizes. We observe
that for all lattice sizes, all the curves intersect each other at a
single point $\beta=6$. This leads to a conclusion that
$T_c/t \sim 0.167$.

  In \Fig{fig:YPScaledPs}(b), we plot the rescaled pair
structure factor versus the universal scaling function $f(L/\xi)$
[see \Eqn{eqn:PSScal}], where $A$ and $T_c$ are chosen such that
all the data points collapse on a single curve, regardless of the
system size. For $A = 0.5$ and $T_c/t = 0.17$ at $\abs{U}/t = 5$,
all data collapse onto a single curve. Hence the estimated
$T_c/t = 0.17$ at $\abs{U}/t = 5$. Similarly, we estimated $T_c$
for various interaction strengths to map out the $T-U$
phase diagram (see \Fig{fig:YPPhaseDiag}).

\vspace{-8mm}
\subsection{Energy Scales}
\label{subsec:YPEScales}
\vspace{-4mm}

\begin{figure}[t]
  \centerline{
  \includegraphics[width=0.96\columnwidth]{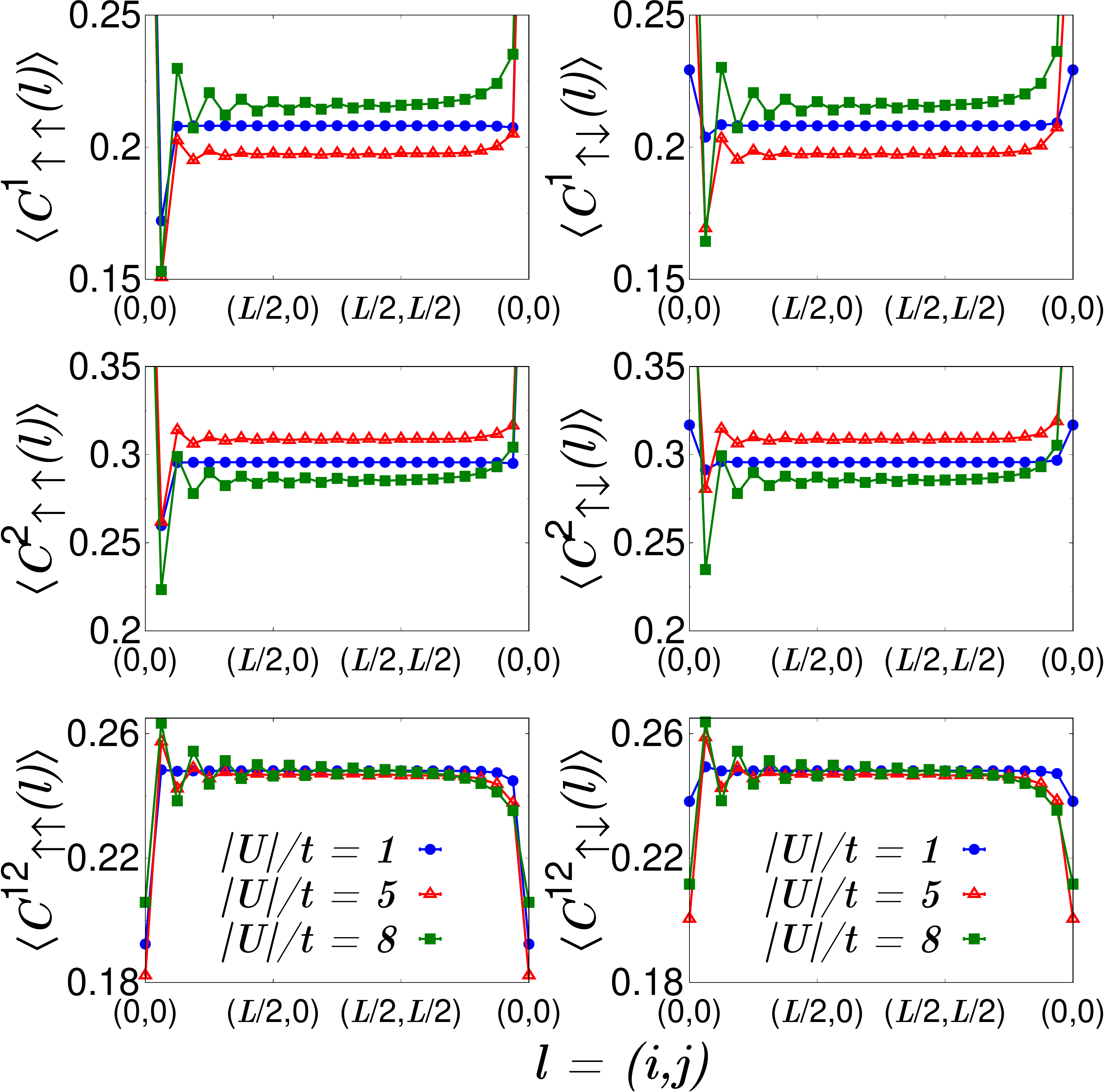}}
  \caption{The spatial variation of the density-density correlation
    function $C_{\sigma,\sigma'}^{\alpha\gamma} ({\bf i,j})$ for
    different interaction strengths $\abs{U}$ at temperature
    $T/t=0.1$ for a system size $N=512$ sites. It indicates the
    formation of the spatial density wave pattern for attractive
    interaction $\abs{U}/t=8$.}
  \label{fig:YPdenslxly}
\end{figure}

  In our bilayer band-insulator model, we see two different energy
scales for the attractive interaction $\abs{U}/t = 5$. One energy
scale $(T^*/t)$ corresponds to the formation of molecules, while the
other energy scale $(T_c/t)$ corresponds to the emergence of the
phase coherence between these pairs. We can identify these two scales
by comparing the evolution of the double occupancy and the $s$-wave
structure factor with temperature. In \Fig{fig:YPdblePsvsU}, we show
the rescaled double occupancy and the $s$-wave structure factor for
$\abs{U}/t=5$ for the lattice size $L=16$. We recover the two energy
scales ($T/t \sim 0.125$ corresponding to saturation of $S_s$ and
$T^*/t \sim 0.33$ corresponding to the saturation of $\tilde{D}$)
and observe the formation of pairs before the emergence of phase
coherence, which is expected in the BEC regime. Finite-size scaling
gives $T_c/t \sim 0.17$.

\vspace{-6mm}
\subsection{Density-Density Correlation}
\label{subsec:YPDensDens}
\vspace{-4mm}

  To study the CDW order, we define the density-density correlation
function as
\vspace{-1.5mm}
\begin{equation}
  \label{eqn:YPCDWdcorr}
  C_{\sigma,\sigma'}^{\alpha \gamma} ({\bf i},{\bf j}) =
  \mean{n_{{\bf i}\sigma}^\alpha n_{{\bf j}\sigma'}^\gamma}
- \mean{n_{{\bf i}\sigma}^\alpha}
  \mean{n_{{\bf j}\sigma'}^\gamma}
\vspace{-1.5mm}
\end{equation}
where $\sigma$ and $\sigma'$ correspond to $\uparrow$ or $\downarrow$
spin, respectively, $\alpha$ and $\gamma$ correspond to layer $A$ or
$B$, respectively, $n_{{\bf i}\sigma}^\alpha$ corresponds to the
fermion density at site $i$ of the $\alpha$th layer, and $i$ and
$j$ run over sites $1$ to $N$. For $\alpha=\gamma$,
$C_{\sigma,\sigma'}^\alpha ({\bf i},{\bf j})$ corresponds to the
intralayer density-density correlation, while for $\alpha\neq\gamma$,
$C_{\sigma,\sigma'}^{\alpha\gamma}({\bf i,j})$ corresponds to the
interlayer density-density correlation function. Similarly, we define
the CDW structure factor $S_{CDW}$ as
\vspace{-1.5mm}
\begin{equation}
  \label{eqn:CDWSF}
  S_{CDW} ({\bf q}) = \frac{1}{N} \sum_{\alpha \gamma}
    \sum_{\sigma\sigma'} \sum_{{\bf ij}}
    e^{{\it i} {\bf q}.{\bf l^{\alpha \gamma}_{{\bf ij}}}} \
    C^{\alpha \gamma}_{\sigma,\sigma'} ({\bf i,j})
\vspace{-1.5mm}
\end{equation}
where ${\bf l^{\alpha\gamma}_{{\bf ij}}} =
{\bf l(i,\alpha) - l(j,\gamma)}$.

  In \Fig{fig:YPdenslxly}, we show the spatial variation of the
density-density correlation function
$C_{\sigma,\sigma'}^{\alpha \gamma}({\bf i,j})$. We observe that the
density wave formation will start to take place for
$\abs{U}/t\geq 8$. Thus there is no competing order in the region
where the superfluid state exists. To confirm this, we perform the
scaling analysis.

\begin{figure}[t]
  \centerline{
  \includegraphics[width=0.96\columnwidth]{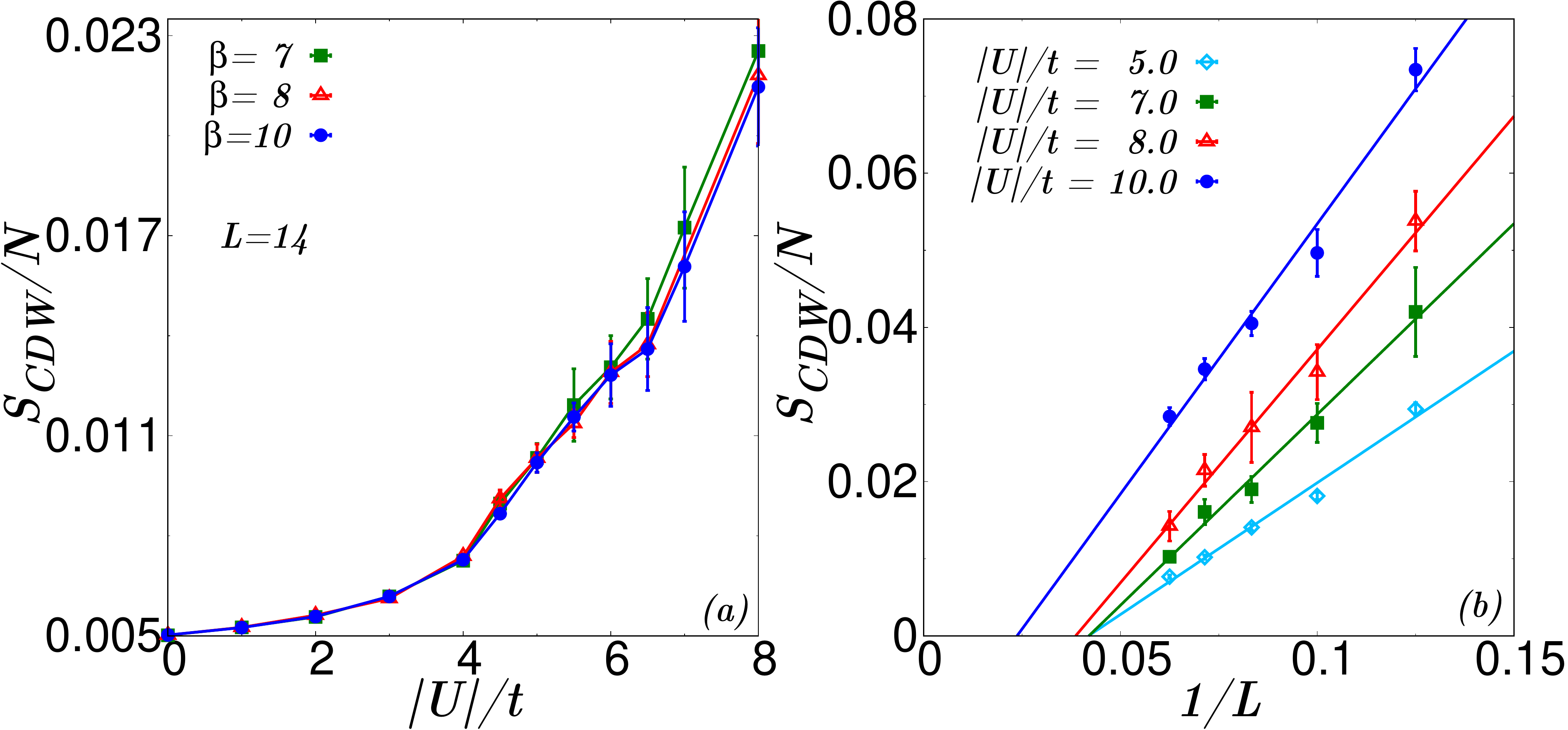}}
  \caption{(a) The evolution of the CDW structure factor $S_{CDW}$
    with the attractive interaction $\abs{U}$. As the attractive
    interaction increases the density wave structure factor
    increases. (b) Finite-size scaling of the CDW structure factor
    for various interaction strengths. It shows the absence of the
    charge density wave upto $\abs{U}/t=10$.}
  \label{fig:YPcdw_fit}
\end{figure}

  In \Fig{fig:YPcdw_fit}(a), we show the evolution of the CDW
  structure factor $S_{CDW}$ with attractive interaction $\abs{U}$
for various temperatures. $S_{CDW}$ increases slowly in the
weak-coupling regime where the system is in a band-insulating state.
As the attractive interaction increases, the density wave structure
factor increases. To investigate the existence of the charge density
wave order in the ground state of the bilayer band-insulating model,
we perform the finite-size scaling.

\vspace{-6mm}
\subsubsection*{{\bf Finite-size Scaling}}
\label{subsubsec:YPFSSD}
\vspace{-4mm}

  Using the Huse's argument of the spin-wave theory, we expect that
the CDW structure factor and the density-density correlation function
behave as
\vspace{-1.5mm}
\begin{equation}
  \label{eqn:YPSCDWDelta}
  \begin{aligned}
    \frac{S_{CDW}}{N} = && \varDelta^{c^2}_0 + \frac{C(U)}{L}, \\
    C(L/2,L/2) = && \varDelta^{c^2}_0 + B(U) L,
  \end{aligned}
\vspace{-1.5mm}
\end{equation}
where $\varDelta^c_0$ is the zero-temperature charge density wave
order parameter and $C$ and $B$ are constants which depend on the
interaction strength $\abs{U}$.

  \Figg{fig:YPcdw_fit}(b) shows the finite-size scaling of the CDW
structure factor $S_{CDW}$ for various interaction strengths. From
\Eqn{eqn:YPSCDWDelta} we see that the intercept along the $y$ axis
gives the square of the zero-temperature CDW order parameter
$\varDelta^c_0$. For the range of attractive interaction strengths
$(5-10)$, the finite-size scaling of the CDW structure factor
confirms that the CDW phase does not exist in the bilayer
band-insulator model. Even though we observe density wave formation
for $\abs{U}/t \ge 8$, to have a long-range CDW order, the CDW
structure factor measured on finite lattices at the critical
temperature $T'_c$ should obey
\vspace{-1.5mm}
\begin{equation}
  \label{eqn:PSScal}
  L^{-7/4} \; S_{CDW} \; \sim \; f(L \ (\beta-\beta'_c)),
\vspace{-1.5mm}
\end{equation}
Hence if we plot $L^{-7/4} S_{CDW}$ as a function of the inverse
temperature $\beta$, different sizes $L$ must cross at
$\beta=\beta'_c$. But we could not see any crossing for different
system sizes for the CDW structure factor. On the contrary, this
crossing has been clearly visible for a pair-pair structure factor
where long-range superfluid order is present (\Fig{fig:YPScaledPs}).
Hence, even for $U/t=10$, long-range CDW order is absent.

%----------------------------------------------------------------------------------------
%       SECTION 4
%----------------------------------------------------------------------------------------

\begin{figure}[t]
  \centerline{
  \includegraphics[width=0.96\columnwidth]{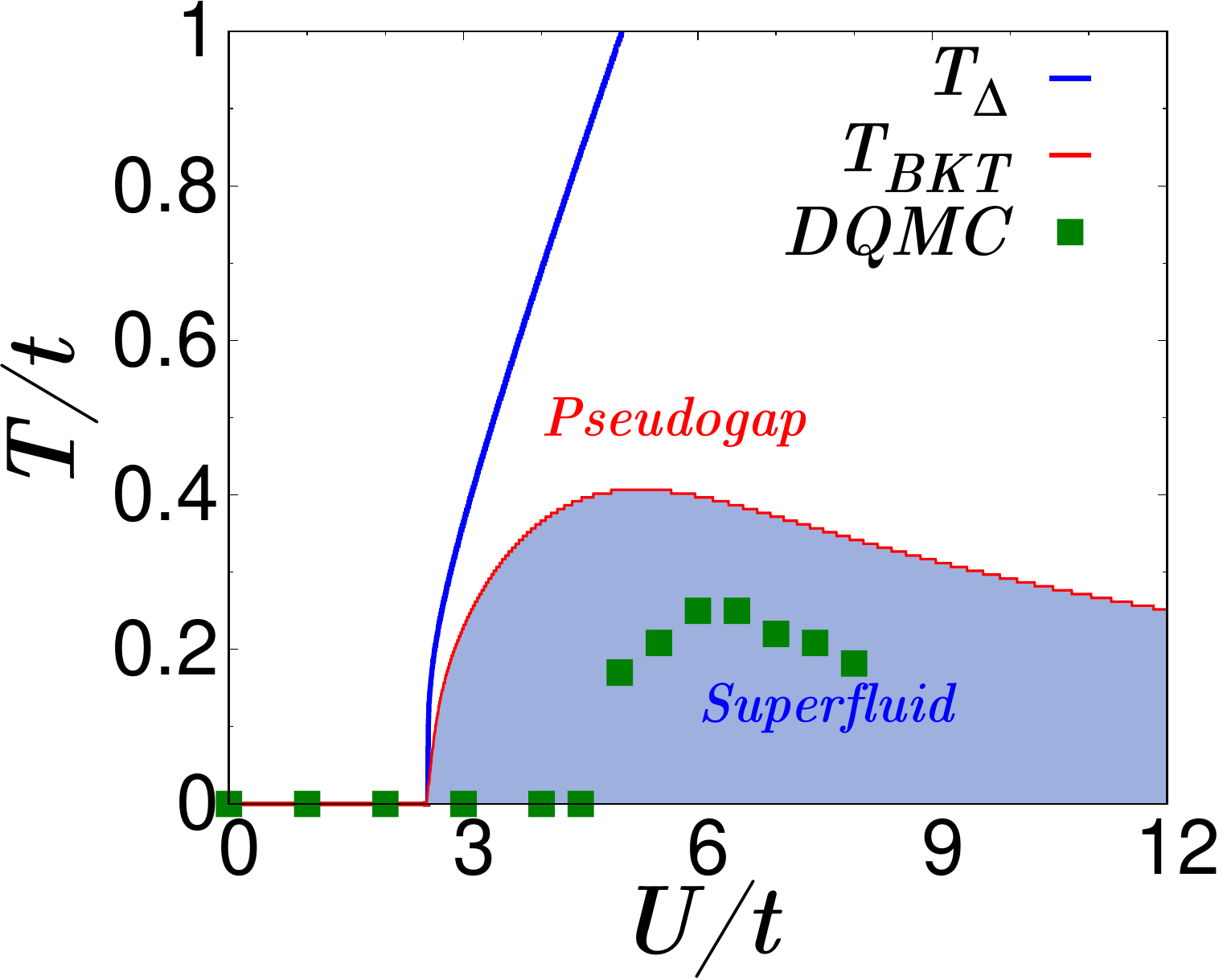}}
  \caption{$T-U$ phase diagram. $T_\Delta$ in the phase diagram
    represents the pair-breaking temperature obtained through
    saddle-point analysis, while $T_{BKT}$ refers to the
    Berezinski-Kosterlitz-Thouless transition temperature of the
    bilayer model presented in \cite{Prasad2014}. Square points
    indicate $T_c$ obtained from the DQMC calculations. We observe
    that the $T_c$ gets reduced from its mean-field values, as
    expected. Also, the maximum value of $T_c(\sim 0.27)$ occurs at
    $\abs{U}/t=6$.}
  \label{fig:YPPhaseDiag}
\end{figure}

\vspace{-6mm}
\section{Discussion and Conclusion}
\label{sec:YPDisConc}
\vspace{-4mm}

  In summary, we have used an unbiased and exact DQMC technique to
study various single-particle and two-particle properties of the
bilayer band-insulator model. We have shown the existence of two
energy scales: one scale governs the phase coherence and the other
one corresponds to the molecule formation (formation of tightly bound
fermionic pairs). We compare our results with the one obtained with
the mean-field and the Gaussian-fluctuation theory results presented
in \cite{Prasad2014}. The critical strength $\abs{U_c}/t\sim 5$ is
slightly higher than the saddle-point analysis $(\abs{U_c}/t=3.2)$
and is close to the VMC prediction $(\abs{U_c}/t=4.5)$. Through the
finite-size scaling, we show that there is no competing CDW order in
the bilayer band-insulator model for the interaction range
$\abs{U}/t=5-10$. The saddle-point analysis suggested the maximum
critical temperature $\sim 0.4$ at $\abs{U}/t=5$, whereas DQMC
predicted $T_c/t|_{DQMC}=0.17$, which is lower than the saddle-point
prediction at $\abs{U}/t=5$. We estimated the transition temperature
for various interaction strengths and map out the $T-U$ phase diagram
shown in \Fig{fig:YPPhaseDiag}. DQMC simulation predicts the maximum
$T_c/t(=0.27)$ which occurs at $\abs{U}/t\sim 6$.
We find that the maximum $T_c$ in our proposed {\it half-filled}
bilayer band insulator is twice that of the maximum $T_c/t\sim 0.13$
(for $\abs{U}/t=8$) published in Ref. \cite{Scalettar1989} for the
single-layer attractive Hubbard model. Thus, the studied bilayer
band-insulator model has ``higher" characteristic temperature $T_c$,
with no competing orders as compared to earlier attempts, and is
expected to be realized in cold-atom experiments.

\vspace{-6mm}
\section*{Acknowledgements}
\vspace{-4mm}

  Y.P. would like to acknowledge CSIR for financial support. Y.P.
thanks A. V. Mallik, A. Halder, and V. B. Shenoy for various
discussions and comments. Y.P. would also like to thank V. B. Shenoy
for the cluster usage.

%----------------------------------------------------------------------------------------

\bibliography{Bibliography.bib}

\end{document}